\documentstyle[epsfig]{aipproc}
\begin{document}

\title{The First Results of K2K long-baseline\\ 
       Neutrino Oscillation Experiment}

\author{Taku Ishida$^*$, representing K2K collaboration}
\address{$^*$Institute for Particle and Nuclear Studies(IPNS)\\
High Energy Accelerator Research Organization(KEK)\\
1-1 Oho, Tsukuba-shi, Ibaraki 305-0801, Japan}

\maketitle

\begin{abstract}
 The first results of the K2K(KEK to Kamioka) long-baseline neutrino
 oscillation experiment are presented in this talk. In 1999
 7.2$\times$10$^{18}$ protons on target were delivered to the
 experiment.  During this period of running there were 3 events fully
 contained in the Super-Kamiokande inner detector fiducial area which
 occurred during the beam spill timing window.  In the case of no
 oscillations the expected number of events during this period
 was 12.3${{+1.7}\atop{-1.9}}$.  The near detectors located at KEK
 also have begun detailed measurements of neutrino
 interactions in water at around 1~GeV.
\end{abstract}

\section*{Introduction}

\begin{figure}[p] 
\centerline{\epsfig{file=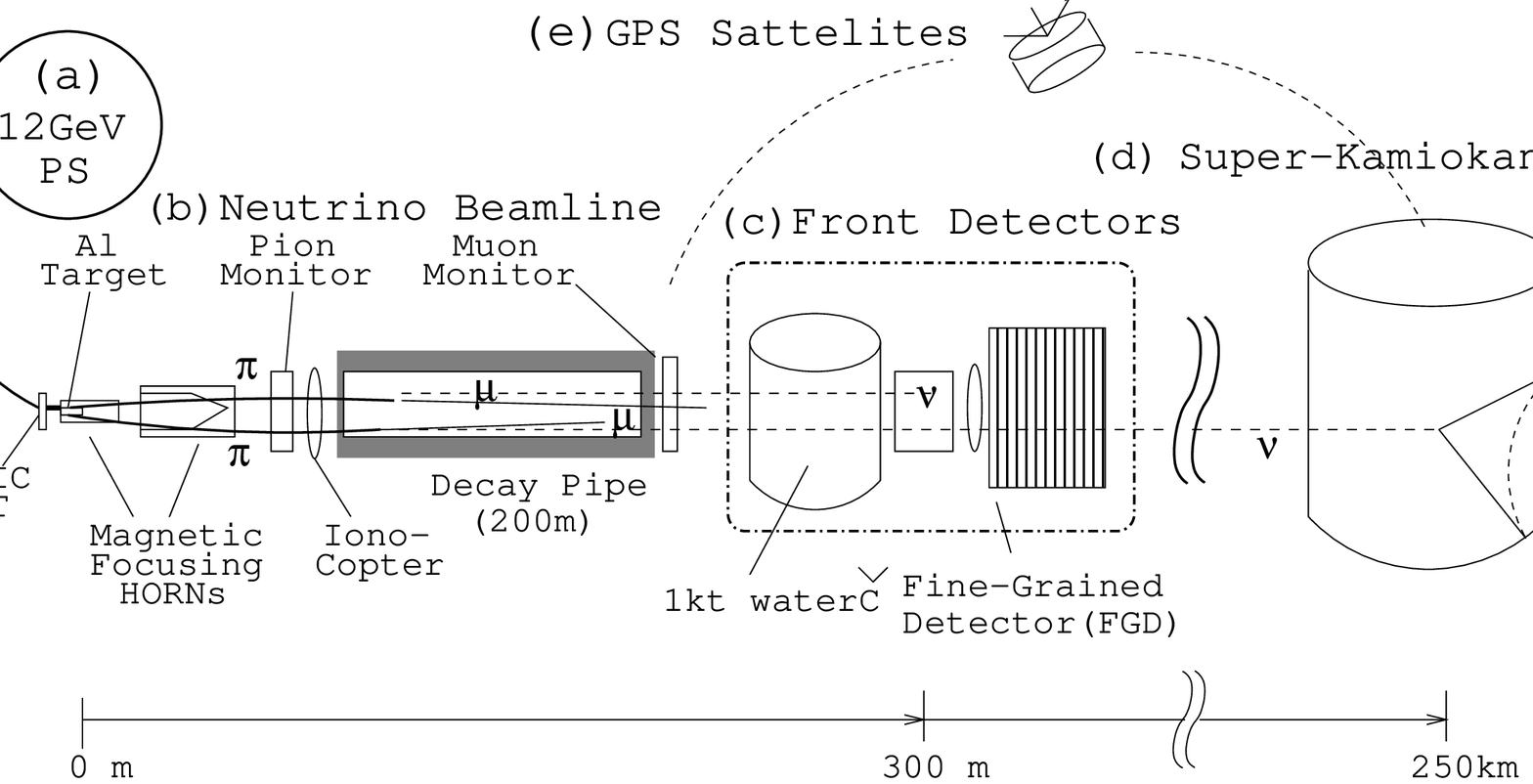,height=2.1in}}
\vspace{10pt}
\caption{Schematic overview of the K2K experiment.
(a) 12 GeV KEK-PS. A spill has 9 bunches in 1.1 $\mu$sec,
$7\times10^{12}$ $ppp$ for every 2.2 sec.  (b) Neutrino beam
line. Protons, bent to the direction of Kamioka in the ``arc section''
of the beam line, are injected into Aluminum target of 3cm$\phi\times
66cm$, embedded in the 1st horn. The proton beam profile and strength
before the target are measured by 2 SPICs and 2 CTs, respectively.
Two horn magnets for $\pi$ focusing, a gas Cherenkov counter for $\pi$
momentum distribution measurement ($\pi$ monitor), and $\phi$-symmetry
monitor (iono-copter) are in a target station.  $\pi$s decay into
$\mu$ and $\nu_\mu$ in 200m of decay pipe filled with helium gas. The
$\mu$ profile is measured by ionization chambers and a silicon pad
detector located behind the beam dump ($\mu$ monitor).  (c) Front
detector system to measure neutrino beam properties at the
production. It is composed of a SK-like water Cherenkov detector
(1$kt$ detector) and a fine-grained detector(FGD) for detailed study
of neutrino interaction with water.  (d) Super-Kamiokande(SK) as a far
detector, 50 $kt$ water Cherenkov detector under stable operation
since 1996, 250 $km$ downstream of the target. (e) The GPS system is
used to look for events at SK during the KEK PS beam pulse. The
precision of $\Delta(T_{\rm KEK}-T_{\rm SK})$ is within 300 $nsec$,
calibrated by an atomic clock.  }
\label{fig1}

\vspace{10pt}

\centerline{\epsfig{file=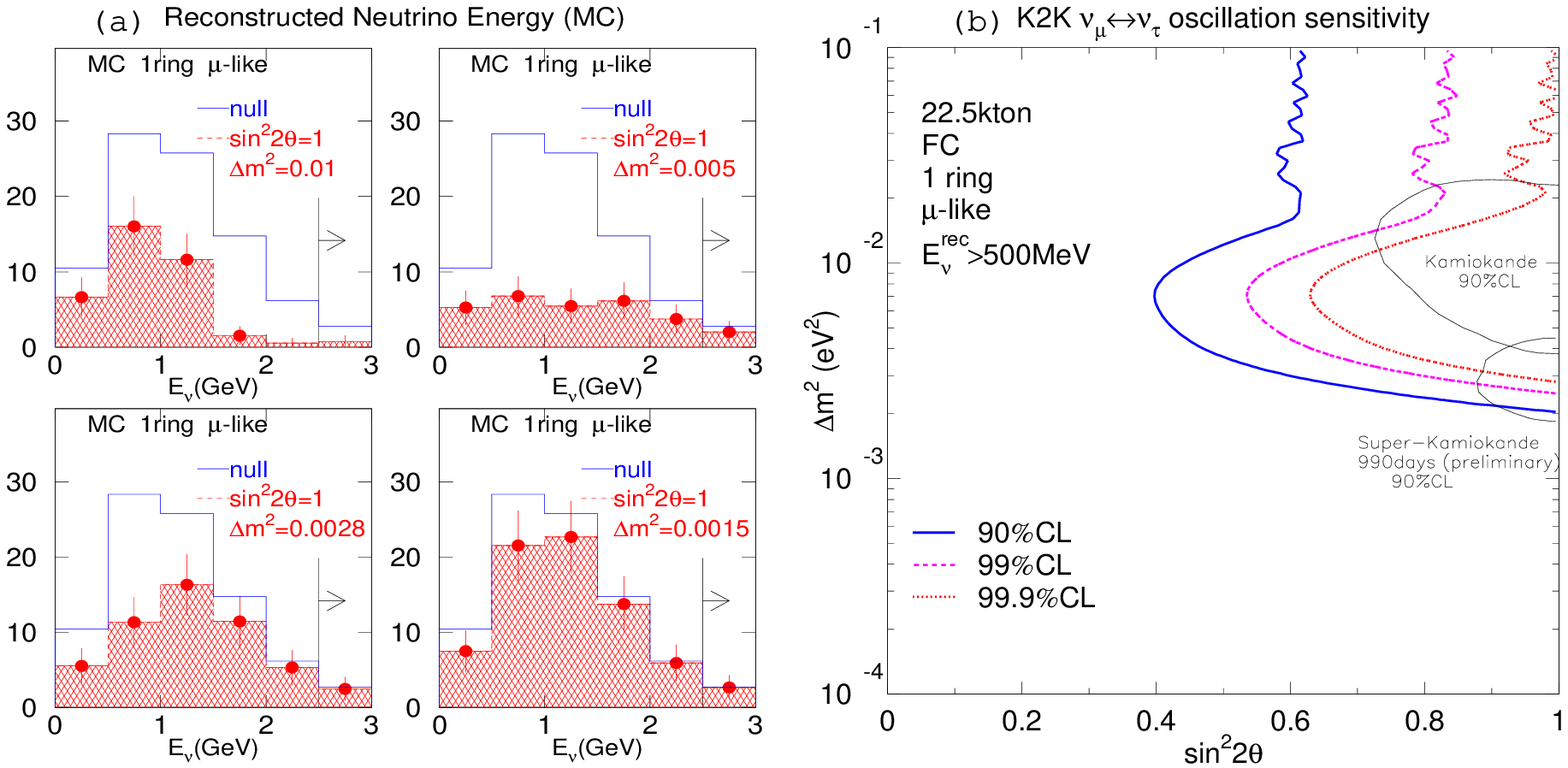,height=2.2in}}
\vspace{10pt}
\caption{(a) Monte Carlo simulated
reconstructed neutrino spectrum at SK, with four
oscillation parameter sets.  10$^{20}$ protons on target correspond to
190 neutrino interactions in the 22.5 $kt$ SK inner fiducial volume in
case of null oscillation. Plots are for 90 fully contained single-ring
$\mu$-like events.  (b) Sensitivity plots of the K2K experiment for
$\nu_\mu\leftrightarrow\nu_\tau$ oscillation.  }
\label{fig2}
\end{figure}

The atmospheric neutrino anomaly observed by Super-Kamiokande(SK) and
other recent underground experiments strongly suggests
$\nu_\mu\leftrightarrow\nu_\tau$ neutrino oscillation.  The allowed
region of the oscillation parameters obtained by SK are in the range
of $\Delta m^2=2\sim 5 \times 10^{-3}$ eV$^2$ and $sin^2(2\theta) >$
0.88 with 90\% confidence level\cite{rf1}, where $\Delta m^2$ is the
mass difference squared between two neutrino mass eigenstates and
$\theta$ is the mixing angle between two neutrinos.

The principal goal of the K2K experiment is to confirm neutrino
oscillation with a man-made neutrino beam and to measure the
oscillation parameters.  Fig.~\ref{fig1} is a schematic of the
setup.  We use the 12 GeV KEK-PS as a neutrino source and
SK as the far detector. The distance between KEK and SK is
250$km$, and the neutrino beam has an average energy of 1.4 GeV.  The
neutrinos are produced by charged pion decays and expected to be 99\%
pure $\nu_\mu$ with an angular deviation $\leq$ 3 $mrad$.  In order to
measure the effects of oscillation we compare the $\nu_\mu$ spectrum
observed by SK to the one measured in the front detectors at the
point of production.

Fig.~\ref{fig2}(a) shows the Monte Carlo simulation of reconstructed
neutrino spectra at SK for $1\times 10^{20}$ protons on target($pot$).
This corresponds to approximately 5 years of measurement.  Oscillated
spectra for four specific oscillation parameter sets are shown.  The
effects of oscillation are seen in the figures as a significant
divergence from the spectra of the null oscillation case, which is
given by open histograms.  K2K's 90\% confidence level
sensitivity covers almost all of the 90\% allowed regions obtained by
Kamiokande and SK, as shown in Fig.~\ref{fig2}(b).

\begin{figure}[t!] 
\centerline{\epsfig{file=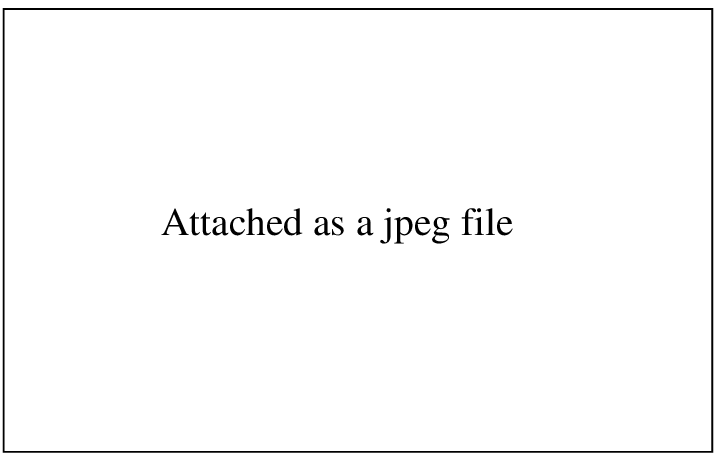,height=2.1in}}
\vspace{10pt}
\caption{(a) Integrated $pot$ and (b) $ppp$ at the target, as functions of
date from April 1999 to March 2000. Dates of SK fully-contained events 
are marked with arrows for 1999.
}
\label{fig3}
\end{figure}

Fig.~\ref{fig3} shows the record of protons on target.
After the completion of the front detector construction in
Feb.~1999, the fast extraction of protons for the experiment started
on Feb. 3. On Mar. 4, the neutrino beam DAQ started with a horn current
of 175 kA and with proton intensity of 3$\times$10$^{12}ppp$. After
engineering runs to study neutrino beam operations in April through
May, stable data taking began in June with an aluminum target with
2$cm\phi$, a horn current of 200 kA, and a proton intensity of
4.5$\times 10^{12} ppp$.  After the summer shutdown, continuous data
taking began again in November, this time with an aluminum target of
$3cm\phi$ and a horn current of 250 kA with 5$\times 10^{12} ppp$. In
1999 we accumulated 7.2 $\times 10^{18} pot$ in total.

\section*{Observation by near detectors}

The near detector system consists of a 1kt water Cherenkov
detector(1kt) and a fine-grained detector(FGD).  The latter consists
of a scintillating fiber tracker(SFT)\cite{rf2}, plastic scintillator
veto counters , an electromagnetic calorimeter of 600 lead glass
blocks, and a muon ranger of 12 iron plates
(10$cm\times$4+20$cm\times$8) with drift chambers(MUC).  The SFT is
composed of 19 layers of 6$cm$-thick water containers sandwiched with
20$\times$($yy$-$xx$) layers of 700$\mu m\phi$ scintillating fibers.

1kt is used for normalizing predicted beam flux at
SK. The fact that the two detectors are so similar cancels out
systematic errors that would otherwise be present.

MUC contained events are used to check the stability of the neutrino
beam.  This is because the large fiducial volume mass of the MUC
results in a very large event rate. The beam center is observed to be
stable to within 1$mrad.$. In addition, spill by spill beam centering,
monitored by the profile of the muons produced by pion-decay that
penetrate the beam dump, was also found to be within 0.5$mrad$.

\begin{figure}[t!] 
\centerline{\epsfig{file=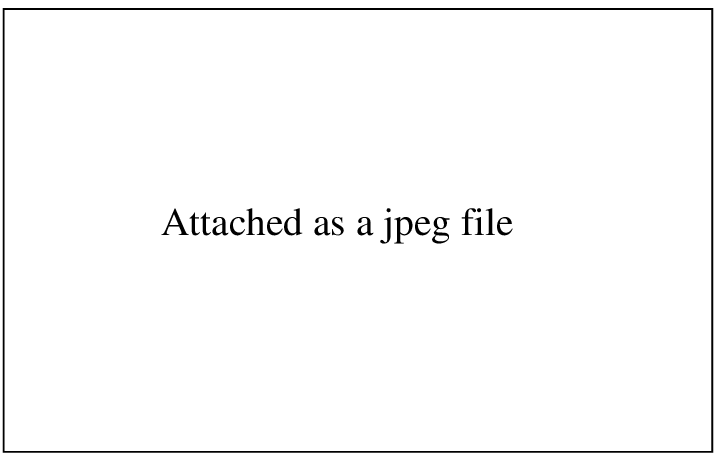,height=1.6in}}
\vspace{10pt}
\caption{(a) A typical $\nu_\mu$ event observed with the FGD. 
(b) Muon energy distribution for single track samples.
(c) $cos(\Delta\theta_{P})$ distribution for 2 track samples.
In (b)/(c), histograms show MC, where open area is for quasi-elastic events,
and hatched area is for non-QE (charged current inelastic + neutral current)
events.}
\label{fig4}
\end{figure}

Fig.~\ref{fig4}(a) is an example of quasi-elastic(QE) neutrino event
candidates with vertex in the SFT. The primary goal of the FGD is to
reconstruct the neutrino spectrum by using QE samples.
(b) shows the reconstructed muon energy distribution
for single track events, and (c) is $cos(\Delta\theta_P)$
distribution for 2 track samples, where $\Delta\theta_P$ is
angular difference between reconstructed proton track and
that calculated from muon momentum. The single track samples and 2
track samples with $cos(\Delta\theta_P)\geq 0.95$ contain a
large fraction of QE events, $\sim$60\% and $\sim$80\% respectively.

We also study the ratio of inelastic events to QE events, in order to
reduce the uncertainty of the calculated neutrino interaction cross
section to less than 10\%. This information is important not only for
K2K, but also for the SK atmospheric neutrino
analyses. Table~\ref{table1} gives summary of the front detector
results.  Event numbers normalized by MC predictions agree very well
to each other.
\begin{table}[b]
\caption{Summary of the near detector observation.}
\label{table1}
\begin{tabular}{lcrcrc}
        ~         & Mass(t)    &       ~  &  $pot$(10$^{18}$) &
 Events           &
 \multicolumn{1}{c}{Data/MC$\pm${\it st.}$\pm${\it sys.}} \\
\tableline
{\bf 1kt}            &  50.3    & June       &    2.03    
                     &   8,157    &   0.83$\pm$.01$\pm$.07   \\
                     &          & November   &    2.62    
                     &  11,337    &   0.84$\pm$.01$\pm$.11   \\ 
{\bf MUC contained}  &  445.    & June       &    3.02    
                     &  27,985    &   0.83$\pm$.01$\pm$.11   \\
                     &          & November   &    2.75    
                     &  28,077    &   0.86$\pm$.01$\pm$.11   \\ 
{\bf SFT+MUC}        &  4.9     & June       &    2.28    
                     &     315    &   0.83$\pm$.05$\pm^{.08}_{.09}$   \\
                     &          & November   &    1.94    
                     &     347    &   0.86$\pm$.05$\pm^{.08}_{.09}$  \\
\end{tabular}
\end{table}

\section*{Neutrino events in Super-Kamiokande}

\begin{figure}[t!] 
\centerline{\epsfig{file=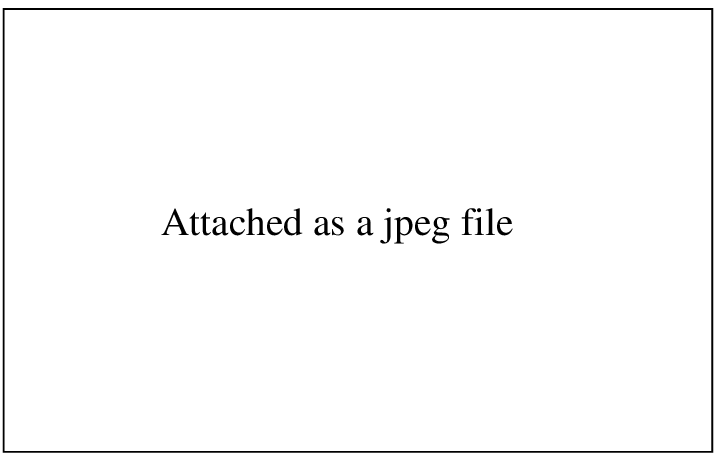,height=1.6in}}
\vspace{10pt}
\caption{SK fully contained event examples  (a) the 1st event in June, 
and (b) in November. (c) $\Delta T$ ($\equiv
T_{SK}-T_{KEK}-T_{T.O.F.}$) distributions at each reduction stage:
$\pm$ 500$\mu sec$ time cut, (1) $\mu$-decay electron cut (2) high
energy trigger condition (3) inner counter total {\it p.e.} cut
$200<Q_{TOT}<50,000 p.e.$, and (4) outer detector cut.    }
\label{fig5}
\end{figure}
On June 19, 1999, 18:42(JST), K2K observed its first neutrino event
due to the KEK neutrino beam in Super-Kamiokande. This was the first
time that an artificially produced particle was detected after
traveling such a large distance. 
Fig.~\ref{fig5}(c) shows  
$\Delta T$ distributions at each reduction step,
where $\Delta T$ is the time difference between SK event and KEK beam pulse,
taking time of flight between KEK and SK into account. 
It is obtained by GPS within the precision of 
300 $nsec$. During the period of running we have 6
events within 1.3$\mu sec$ time window, 3 of which are within 22.5$kt$
fiducial volume (vertex distance from wall $\geq$ 2$m$).

We can estimate the number ($N_{SK}^{pred}$)
using the observed number of events at 1kt($N_{KEK}$): $ N_{SK}^{pred}=
(N_{KEK} / \epsilon_{KEK}) \cdot R \cdot \epsilon_{SK}$, where $R$ is
a factor of extrapolation from KEK to SK, $\epsilon_{KEK}$ and
$\epsilon_{SK}$ are the detection efficiency of 1kt and SK, respectively.  
Since the target material (water) is common and the
systematic uncertainty due to the cross section cancels, the value
of R depends mostly on the flux ratio between SK and KEK, whose
uncertainty is ${{+8\%}\atop{-10\%}}$. 
$N_{SK}^{pred}$ was estimated to be 12.3${{+1.7}\atop{-1.9}}$
in total.
Table~\ref{table2} summarizes these numbers, with 
expectations for the case of three typical
$\Delta m^2$ with $sin^2(2\theta)=1$. We
will accumulate $>2\times 10^{19}pot$ in this summer, leading to an
increase in the statistical power of these results.

\begin{table}[b]
    \caption{Summary of SK events. Fully-Contained(FC) are events with 
the significant light detected in the inner detector only. 
Outer Detector(OD) events, with light
detected in the outer detector is also tabulated for reference 
(systematic uncertainty$\sim$ 40\%).}
    \label{table2}
\begin{tabular}{lccccc}
 & Data &   No    oscillation  & $\Delta m^2=3\times 10^{-3}$   & 
$5\times 10^{-3}$ &   $7\times
10^{-3}\ (eV)^2$   \\   
\tableline
FC in fiducial& 3 & 12.3\hbox{${ {+1.7}\atop{-1.9} }$}  & 8.0     & 5.4   
 
 & 4.6  \\
out of fiducial   & 3 & 5.5\hbox{${{+1.1}\atop{-1.2}}$} & 3.5  & 2.4    & 
2.1 \\
\tableline
OD contained     & 4& $8.7\pm 3.3$   & 5.5     & 3.5     & 2.9  \\
(inner crossing) & 2 & $4.2\pm 1.6$ & 3.2     & 2.0    & 1.3 \\
\end{tabular}
\end{table}
\end{document}